\documentclass[a4paper,11pt]{article}
\usepackage{pos}
\usepackage{bm}
\usepackage{braket}
\usepackage{slashed}

\title{\textit{Ab initio} nuclear thermodynamics from  lattice effective field theory}

\author*[a]{Bing-Nan Lu}
\author[b]{Ning Li}
\author[c]{Serdar Elhatisari}
\author[d]{Dean Lee}
\author[e]{Joaqu{\'i}n~E.~Drut}

\author[f]{Timo A. L\"ahde}

\author[g]{Evgeny Epelbaum}

\author[h, i, j]{Ulf-G. Mei{\ss}ner}

\affiliation[a]{Graduate School of China Academy of Engineering Physics, Beijing
100193, China}

\affiliation[b]{School of Physics, Sun Yat-Sen University, Guangzhou 510275, China}

\affiliation[c]{Faculty of Natural Sciences and Engineering, Gaziantep Islam Science and Technology University, Gaziantep 27010, Turkey}

\affiliation[d]{ Facility for Rare Isotope Beams and Department of Physics and Astronomy, Michigan State University, MI 48824, USA}

\affiliation[e]{Department of Physics and Astronomy, University of North Carolina,
Chapel Hill, North Carolina 27599-3255, USA }

\affiliation[f]{Institute for Advanced Simulation, Institut f\"ur Kernphysik, Center for Advanced Simulation and Analytics, and
J\"ulich Center for Hadron Physics, Forschungszentrum J\"ulich,
D-52425 J\"ulich, Germany}

\affiliation[g]{Ruhr-Universit{\"a}t Bochum, Fakult{\"a}t f{\"u}r Physik und Astronomie,
Institut f{\"u}r Theoretische Physik II, D-44780 Bochum, Germany}

\affiliation[h]{Helmholtz-Institut f\"ur Strahlen- und Kernphysik and Bethe Center
for Theoretical Physics, Universit\"at Bonn, D-53115 Bonn, Germany}

\affiliation[i]{Institute for Advanced Simulation, Institut f\"ur Kernphysik, and
J\"ulich Center for Hadron Physics, Forschungszentrum J\"ulich,
D-52425 J\"ulich, Germany}

\affiliation[j]{Tbilisi State University, 0186 Tbilisi, Georgia}

\emailAdd{bnlv@gscaep.ac.cn}

\abstract{We show that the \textit{ab initio} calculations of nuclear thermodynamics can be performed efficiently
using lattice effective field theory.
The simulations use a new approach called the pinhole trace algorithm to calculate
thermodynamic observables for a fixed number of protons and neutrons enclosed in a
finite box.
In this framework, we calculate the equation of state, the liquid-vapor coexistence line and the critical point of neutral
symmetric nuclear matter with high precision.
Since the algorithm uses a
canonical ensemble with a fixed number of particles, it provides a sizable computational advantage over
grand canonical ensemble simulations that can be a factor of several thousands to as much as
several millions for large volume simulations.}

\FullConference{%
 The 38th International Symposium on Lattice Field Theory, LATTICE2021
  26th-30th July, 2021
  Zoom/Gather@Massachusetts Institute of Technology
}

\begin{document}
\maketitle

\section{Introduction}
\begin{figure}
\begin{centering}
\includegraphics[width=0.6\columnwidth]{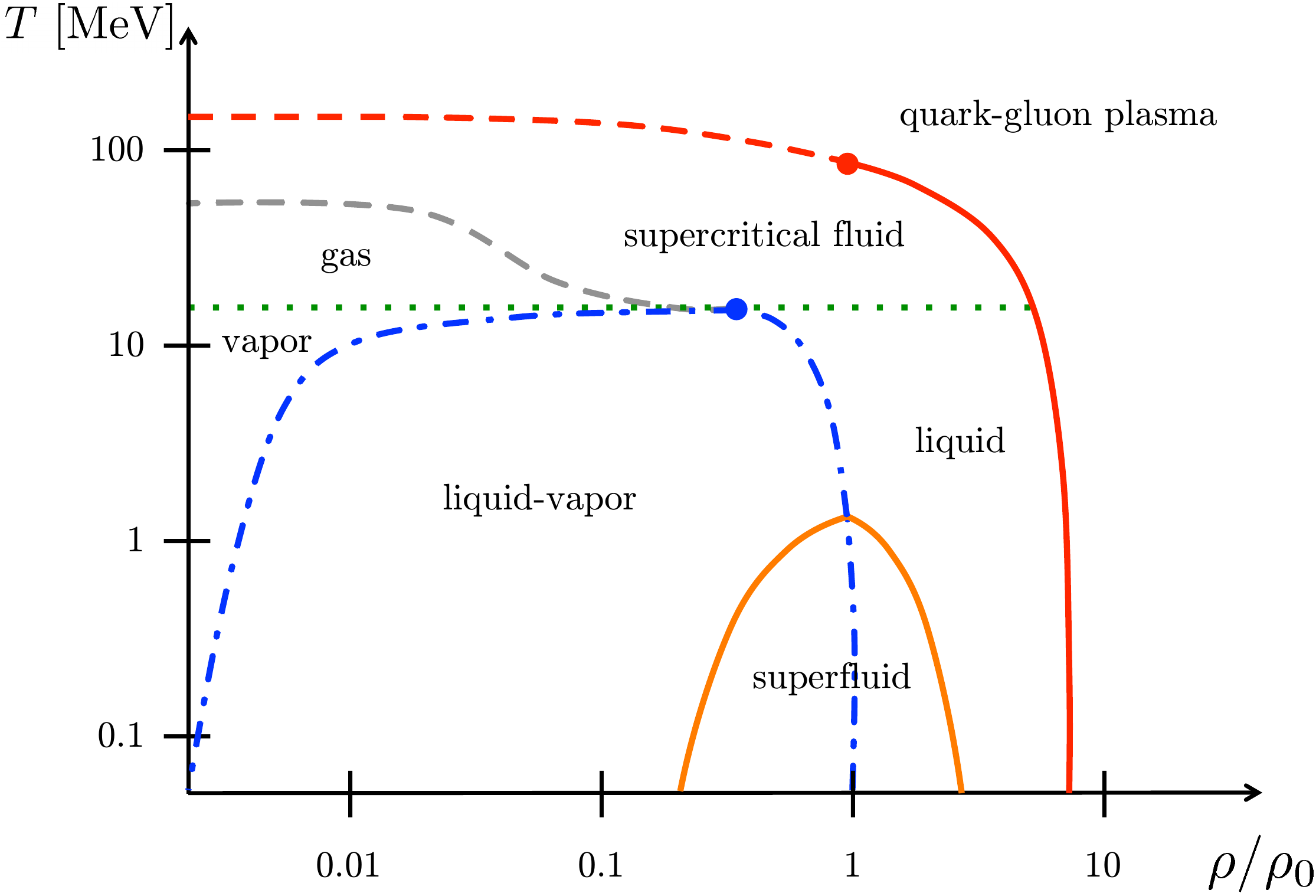}
\par\end{centering}
\caption{\label{fig:phase_diagram}Phase diagram of symmetric nuclear matter.  The horizontal axis
is nucleon density $\rho$ as a fraction of saturation density $\rho_0$, and the vertical axis is temperature in units of MeV.  The solid lines indicate first-order transitions, the filled circles are critical points, and the dashed lines indicate the approximate locations where crossover transitions occur.  The dashed-dotted line outlines the liquid-vapor coexistence region, with the entire region marking the first-order liquid-vapor transition.  The dotted line marks the critical temperature for the liquid-gas critical point.}
\end{figure}

The equation of state of strongly interacting matter is one of the central topics in
contemporary nuclear physics, as it plays an important role in the early universe,
heavy-ion reactions and the generation of gravitational waves in violent neutron star mergers.
In Fig.~\ref{fig:phase_diagram} we show
the phase diagram of symmetric nuclear matter with equal numbers of protons and neutrons (or equal numbers of up and down quarks).
The horizontal axis is the nucleon density $\rho$ as a fraction of the saturation density $\rho_0$, and
the vertical axis is temperature in units of MeV.  The nuclear equation of state for both symmetric and
asymmetric matter is of great relevance to the evolution and dynamics of
core-collapse supernovae \cite{Togashi2017}, neutron star cooling \cite{Page2004}, and neutron star mergers \cite{Most2019}.  There are also important connections between
the nuclear equation of state and heavy-ion collisions.  It is well
established that highly-excited nuclear states can be treated {\it en masse} as part of an equilibrium thermal distribution. The large density of states at
high energies allows a treatment
in terms of thermodynamic concepts, such as temperature,
entropy, and free energy. Simple statistical models
have been used to address processes such as compound nucleus reactions \cite{Hauser1952},
nuclear multifragmentation \cite{Bohr1936},
nuclear liquid-gas phase transitions \cite{Siemens1983}, and stellar nucleosynthesis \cite{Seeger1965}.  Hot nuclei can, for example, be modeled by a simple Fermi gas model (FGM), where many-body correlations and shell
effects are neglected.

Although it can explain some basic properties, the FGM
does not reflect many details of nuclear structure and fails in
explaining phenomena associated with strong correlations such as clustering \cite{Freer2018}.
At relatively low temperatures, these problems have been solved
by many-body methods such as the shell model Monte Carlo (SMMC) approach \cite{Koonin1997},
which includes many-body correlations within a major shell using
stochastic methods. When applied to medium mass nuclei, the SMMC method
improves the FGM level densities for excitation energies of a few
MeV, which is important for both slow and rapid neutron capture processes in astrophysics \cite{Dean1995,Nakada1997,Alhassid2000}.
At higher temperatures, however, continuum states comprised of nucleons and nuclear fragments play important roles in nuclear breakup, and these are not well described in the shell model valence space. In
such cases, the available methods are transport models \cite{Aichelin1991,Ono1992b}, cluster models \cite{Fisher1967,Fisher1967b}, and molecular dynamics~\cite{Furuta2006}, where clustering
correlations are either disregarded or included explicitly. Recent
efforts to introduce important correlations into transport models
can be found in Ref.~\cite{Ono2019}. As each of the above
methods employs effective interactions selected to reproduce a few chosen observables, the uncertainties can be significant, especially at high temperatures or densities where less empirical data is available.

In recent years much progress has been made in \textit{ab
initio} or fully microscopic calculations of the nuclear Hamiltonian starting from underlying nuclear forces. 
When combined with a systematic framework for the nuclear forces such as effective field theory \cite{Epelbaum2009}, these first-principles calculations can reduce systematic errors order by order
and generate important many-body correlations such as those responsible for clustering. 
Unfortunately, most \textit{ab initio} methods rely on computational strategies which are not designed for calculations at nonzero temperature. One exception is the method of lattice effective field theory.  There have been some early efforts to describe nuclear thermodynamics using lattice simulations \cite{Muller1999,Lee2004}, however there has been little progress to report since then.  The difficulties stem from the amount of computational effort needed to perform grand-canonical simulations of nucleons in large spatial volumes.

Recently we developed a new paradigm for calculating \textit{ab
initio} nuclear thermodynamics with lattice simulations~\cite{Lu2020}.
We demonstrated an
efficient method, called the pinhole trace algorithm, for computing nuclear
observables at nonzero temperature using a canonical ensemble with fixed numbers
of protons and neutrons.
In the following, we discuss the formalism in more detail.


\section{Method}

\subsection{Lattice Hamiltonian\label{subsec:LatHam}}

Our \textit{ab initio} calculations are based on nuclear lattice
effective field theory (NLEFT) using a leading-order
pionless EFT interaction as defined in \cite{Lu2018}.  Despite the simplicity of the
interaction, the ground-state energies and charge radii of
the light and medium-mass nuclei are well reproduced, as well as the zero-temperature
equation of state of pure neutron matter \cite{Lu2018}.
When applied to zero-temperature neutral symmetric nuclear matter, we obtain $\rho_{0} = 0.205$~fm$^{-3}$ and $E/A = -16.9$~MeV at the saturation point.
The lattice simulations are performed using auxiliary-field Monte Carlo as described in the review~\cite{Lee2009} and the book~\cite{Lahde2019}.

On a periodic $L^3$ cube with lattice coordinates $\bm{n}=(n_{x,}n_{y},n_{z})$, The Hamiltonian is
\begin{equation}
H_{{\rm SU(4)}}=K+\frac{1}{2!}C_{2}\sum_{\bm{n}}:\tilde{\rho}^2(\bm{n}):+\frac{1}{3!}C_{3}\sum_{\bm{n}}:\tilde{\rho}^3(\bm{n}):,\label{eq:HSU4}
\end{equation}
where $K = -\nabla^2 / (2m)$ is the kinetic energy term with nucleon mass
$m=938.9$~MeV and the $::$ symbol indicate normal ordering.
 The density operator $\tilde{\rho}(\bm{n})$ is defined as
\begin{equation}
\tilde{\rho}(\bm{n})=\sum_{i}\tilde{a}_{i}^{\dagger}(\bm{n})\tilde{a}_{i}(\bm{n})+s_{L}\sum_{|\bm{n}^{\prime}-\bm{n}|=1}\sum_{i}\tilde{a}_{i}^{\dagger}(\bm{n}^{\prime})\tilde{a}_{i}(\bm{n}^{\prime}),
\label{eq:smeared_density}
\end{equation}
with $i$ the joint spin-isospin index, the smeared annihilation and creation operators are defined as
\begin{equation}
\tilde{a}_{i}(\bm{n})=a_{i}(\bm{n})+s_{NL}\sum_{|\bm{n}^{\prime}-\bm{n}|=1}a_{i}(\bm{n}^{\prime}).
\end{equation}
The summation over the spin and isospin implies that the interaction is SU(4) invariant. The parameter $s_L$ controls the strength of the local part of the interaction, while $s_{NL}$ controls the strength of the nonlocal part of
the interaction.
Here we include both kinds of smearing.  Both $s_L$ and $s_{NL}$ have an impact on the range of the interactions.  The parameters $C_{2}$ and $C_{3}$ give the strength of the two-body
and three-body interactions, respectively.

In this work we use a lattice spacing $a=1.32$~fm, which corresponds
to a momentum cutoff $\Lambda=\pi/a\approx471$~MeV.
For the SU(4) interaction we use the parameter set $C_2 = -3.41\times 10^{-7}$ MeV$^{-2}$,
$C_3 = -1.4 \times 10^{-14}$ MeV$^{-5}$, $s_{L} = 0.061$ and $s_{NL} = 0.5$.
These parameters are adjusted to reproduce the deuteron, triton and the properties of medium mass nuclei.

We use a discrete auxiliary field that can simulate
the two-, three- and four-body forces simultaneously without sign oscillations. This follows
from the exact operator identity,
\begin{equation}
:\exp\left(-\frac{1}{2}C\rho^{2}-\frac{1}{6}C_{3}\rho^{3}-\frac{1}{24}C_{4}\rho^{4}\right):=\sum_{k=1}^{N}\omega_{k}:\exp\left(\sqrt{-C}\phi_{k}\rho\right):\label{eq:Auxiliary_Field_Identity}
\end{equation}
where $\rho$ is the one-particle density, $C$ is the interaction coefficient, $C_{3}$ and $C_{4}$ are
coupling constants for three-body and four-body forces, respectively,
$\omega_{k}$'s and $\phi_{k}$'s are real numbers.
In this
work we only consider attractive two-body interactions with $C<0$.
In order to avoid the sign problem we further require $\omega_{k}>0$
for all $k$.

To determine the constants $\phi_{k}$'s and $\omega_{k}$'s, we expand
Eq.~(\ref{eq:Auxiliary_Field_Identity}) up to $\mathcal{O}(\rho^{4})$
and compare both sides order by order.
Here we use the following ansatz,
\begin{equation}
\omega_{1}=\frac{1}{\phi_{1}(\phi_{1}-\phi_{3})},\qquad\omega_{2}=1+\frac{1}{\phi_{1}\phi_{3}},\qquad\omega_{3}=\frac{1}{\phi_{3}(\phi_{3}-\phi_{1})}\label{eq:solutions}
\end{equation}
where $\phi_{2}=0$ and $\phi_{1}$ and $\phi_{3}$ are two
roots of the quadratic equation,
\begin{equation}
\phi^{2}+\frac{C_{3}}{\sqrt{-C^{3}}}\phi-\frac{C_{3}^{2}}{C^{3}}+(\frac{C_{4}}{C^{2}}-3)=0.\label{eq:N3 value-1}
\end{equation}
Using Vieta's formulas, it is straightforward to verify that Eq.~(\ref{eq:solutions})
satisfies Eq.~(\ref{eq:Auxiliary_Field_Identity}) up to $\mathcal{O}(\rho^{4})$.
For a pure two-body interaction $C_{3,4}=0$, the solution is simplified
to $\phi_{1}=-\phi_{3}=\sqrt{3}$, $\phi_{2}=0$, $\omega_{1}=\omega_{3}=1/6$,
$\omega_{2}=2/3$.
The
corresponding auxiliary field $s(n_{t},\bm{n})$ only assumes three
different values $\phi_{1}$, $\phi_{2}$ and $\phi_{3}$ and can
be sampled with the shuttle algorithm described below.

\subsection{Pinhole trace algorithm}

For a canonical ensemble with fixed nucleon number $A$, volume $V$
and temperature $T$, the expectation value of any observable $\mathcal{O}$
can be measured as
\begin{equation}
\langle\mathcal{O}\rangle_{\beta}=\frac{Z_{\mathcal{O}}(\beta)}{Z(\beta)}=\frac{{\rm Tr}_{A}(e^{-\beta H}\mathcal{O})}{{\rm Tr}_{A}(e^{-\beta H})},\label{eq:thermal_average}
\end{equation}
where $Z(\beta)$ is the canonical partition function, $\beta=T^{-1}$ is the inverse temperature, $H$ is the Hamiltonian, and ${\rm Tr}_{A}$ is the trace over the $A$-body Hilbert
space.
Throughout, we work in canonical units with $\hbar = c = k_B = 1$.
In this work
we use a novel algorithm called pinhole trace algorithm (PTA)
to efficiently compute $Z(\beta)$ and $Z_{\mathcal{O}}(\beta)$ on
the lattice. The pinhole trace algorithm is an extension of the pinhole algorithm introduced in Ref.~\cite{Elhatisari2017} to sample the spatial positions and spin/isospin indices of the nucleons. The new
feature is that we also perform a quantum mechanical trace over all possible states.

The canonical partition function $Z(\beta)$ can be written explicitly in the single particle basis as
\begin{eqnarray}
   Z(\beta)& = &\sum_{c_1, \cdots , c_A} \langle c_1, \cdots , c_A | \exp(-\beta H)
     |  c_1, \cdots, c_A \rangle, \label{eq:partitionfunction}
\end{eqnarray}
where the basis states are Slater determinants composed of point particles, $c_i = ({\bm{n}_i, \sigma_i, \tau_i})$ are the quantum numbers of the $i$-th particle,
$\sigma_i$ is the spin and $\tau_i$ is the isospin.
On the lattice, the components of $\bm{n_i}$ take integer values from 0 to $L - 1$, where $L$ is the box length in units of the lattice spacing.
The neutron number $N$ and proton number $Z$ are separately conserved, and the summation in Eq.~(\ref{eq:partitionfunction}) is limited to the subspace with the specified values for $N$ and $Z$.



By decomposing the interactions in $H$ using the auxiliary fields, we obtain the path-integral expression for Eq.~(\ref{eq:partitionfunction})
\begin{eqnarray}
    Z(\beta) &=& \sum_{c_1, \cdots, c_A} \int \mathcal{D}s_{1} \cdots \mathcal{D} s_{L_t} \langle c_1, \cdots, c_A | \times \\
             & & M(s_{L_t}) \cdots M(s_1) |  c_1, \cdots, c_A \rangle, \label{eq:partitionfunction_pathintegral}
\end{eqnarray}
where
\begin{equation}
    M(s_{n_t}) = : \exp\left[-a_t K + \sqrt{-a_t C} \sum_{{\bm n}} s_{n_t}({\bm n}) \rho({\bm n})\right] :
    \label{eq:transfermatrix}
\end{equation}
is the normal-ordered transfer matrix for time step $n_t$, and $s_{n_t}$ is our shorthand for all auxiliary fields at that time step \cite{Lee2009,Lahde2019}.
Note that for the leading order interaction proposed in Sec.~\ref{subsec:LatHam}, we need to use summations over indices instead of the path integral over real variables, and the density $\rho(\bm{n})$ should be substituted by the smeared density Eq.~(\ref{eq:smeared_density}).
For a given configuration $s_{n_t}$, the transfer matrix $M(s_{n_t})$ consists of a string of one-body operators
which are directly applied to each single-particle wave function in the Slater determinant.
For notational convenience, we will use the abbreviations $\vec{c} = \{c_1, \cdots, c_A\}$ and $\vec{s} = \{s_1, \cdots, s_{L_t}\}$.


The pinhole trace algorithm (PTA) was inspired by the pinhole algorithm used to sample the spatial positions and spin/isospin of the nucleons \cite{Elhatisari2017}.
However, the purpose, implementation, and underlying physics of the PTA for nuclear thermodynamics are vastly different from the original pinhole algorithm used for density distributions. In the PTA we evaluate Eq.~(\ref{eq:partitionfunction_pathintegral}) using Monte Carlo methods, i.e. importance sampling is used to generate an ensemble $\Omega$ of $\{\vec{s}, \vec{c}\}$ of configurations according to the relative probability distribution
\begin{equation}
    P (\vec{s}, \vec{c}) = \left| \braket{\vec{c}|  M(s_{L_t}) \cdots M(s_1) | \vec{c}} \right|.  \label{eq:PTAprob}
\end{equation}
The expectation value of any operator $\hat {O}$ can be expressed as
\begin{equation}
    \langle \hat{O} \rangle = \langle \mathcal{M}_O (\vec{s}, \vec{c}) \rangle_\Omega / \langle \mathcal{M}_1 (\vec{s}, \vec{c}) \rangle_\Omega,
    \label{eq:measurement}
\end{equation}
where
\begin{eqnarray}
   \mathcal{M}_O(\vec{s}, \vec{c}) &=& \langle\vec{c}| M(s_{L_t}) \cdots M(s_{L_t/2+1})\hat{O} \times \\
                                   & & M(s_{L_t/2}) \cdots M(s_1)| \vec{c} \rangle / P (\vec{s}, \vec{c}).
\end{eqnarray}

To generate the ensemble $\Omega$ we use the Metropolis algorithm to update $\vec{s}$ and $\vec{c}$ alternately.
We first fix the nucleon configuration $\vec{c}$ and update the auxiliary fields $\vec{s}$. Starting from the rightmost time slice $s_1$, we update $s_1, \cdots, s_{L_t}$ successively
 using a shuttle algorithm,
which works as follows.
(1) Choose one time slice $n_{t}$, record
the corresponding auxiliary field as $s_{{\rm old}}(n_{t},\bm{n}$).
(2) Update the corresponding auxiliary fields at each lattice site
$\bm{n}$ according to the probablity distribution $P\left[s_{{\rm new}}(n_{t},\bm{n})=\phi_{k}\right]=\omega_{k}$,
$k=1,2,3$. Note that $\omega_{1}+\omega_{2}+\omega_{3}=1$. (3) Calculate
the determinant of the $A \times A$ correlation matrix $M_{ij}$ using $s_{{\rm old}}(n_{t},\bm{n})$
and $s_{{\rm new}}(n_{t},\bm{n})$, respectively. (4) Generate a random
number $r\in[0,1)$ and make the ``Metropolis test'': if
\[
\left|\frac{\det\left[M_{ij}\left(s_{{\rm new}}(n_{t},\bm{n})\right)\right]}{\det\left[M_{ij}\left(s_{{\rm old}}(n_{t},\bm{n})\right)\right]}\right|>r,
\]
accept the new configuration $s_{{\rm new}}(n_{t},\bm{n})$ and update
the wave functions accordingly, otherwise keep $s_{{\rm old}}(n_{t},\bm{n})$.
(5) Proceed to the neighboring time slice, repeat steps 1)-4), and turn round at the ends of the time
series.

The shuttle algorithm is well suited for small $a_{t}$. In this
case the number of time slices is large and the impact of a single
update is small. As the new configuation is close to the
old one, the acceptance rate is high.
We compared the
results with the HMC algorithm and found that the new algorithm is
more efficient. In most cases the number of independent configurations
per hour generated by the shuttle algorithm is usually three or four
times larger than that generated by the HMC algorithm.

 After updating $\vec{s}$, we then update the nucleon configuration $\vec{c}$.
 To that end, we randomly choose a nucleon $i$ and move it to one of its neighboring sites
 \begin{equation}
   c_i = \{\bm{n}_i, \sigma_i, \tau_i\} \rightarrow c'_i = \{\bm{n}'_i, \sigma_i, \tau_i\},
  \end{equation}
  or flip its spin,
   \begin{equation}
   c_i = \{\bm{n}_i, \sigma_i, \tau_i\} \rightarrow c'_i =  \{\bm{n}_i, -\sigma_i, \tau_i\}.
  \end{equation}
 The corresponding new nucleon configuration $\vec{c}\,'$ is accepted if
 \begin{equation}
     P(\vec{s}, \vec{c}\,') / P(\vec{s}, \vec{c}) > r'  \label{eq:cUpdate_Metropolis}
 \end{equation}
 with $0 \leq r' < 1$ a random number.
 Because in the $\vec{c}$ update only one nucleon is moved or spin flipped at a time, the successive configurations are correlated.
 Only when all nucleons have been updated do we obtain statistically independent configurations.
For calculations described here, we found that about 16 $\vec{c}$ updates for every $\vec{s}$ update
  produced the optimal sampling efficiency.
Next we discuss the computational scaling of our pinhole trace algorithm (PTA) simulations and the comparison with grand canonical simulations based on the well-known BSS method first described in Ref.~\cite{Blankenbecler:1981jt}.  Both are determinant Monte Carlo algorithms for lattice simulations with auxiliary fields, and so the comparison is relatively straightforward.  We consider a system with $A$ nucleons, $V=L^3$ spatial lattice points, and $L_t$ time steps.  We will drop constant factors such as the factor of 4 associated with the number of nucleon degrees of freedom. In the PTA we compute $A$ single nucleon wave functions, where each wave function has $V$ components, and the correlation matrix $M_{ij}$ will be an $A\times A$ matrix.  Meanwhile the BSS algorithm requires computing $V$ single nucleon wave functions, where each wave function has $V$ components, and the correlation matrix $M_{ij}$ will be a $V\times V$ matrix.

For both algorithms, we update the auxiliary fields sequentially according to time step.  We update all of the auxiliary fields at one time step before moving on to the next time step. We now consider a full sweep that updates all of the auxiliary fields.  During this sweep through the auxiliary fields, the cost associated with updating the single nucleon wave functions is $AVL_t$ for the PTA and $V^2L_t$ for the BSS algorithm.

For small $A$ we can update all of the $V$ auxiliary fields for a given time step in parallel.  But as $A$ becomes large, we need to perform $A$ separate updates per time step, with $V/A$ auxiliary fields updated at a time.  For the PTA, the cost of calculating correlation matrices for the full update over auxiliary fields is $A^2VL_t$, and the cost of calculating matrix determinants for the full update is $A^4L_t$.  For the BSS algorithm, the cost of calculating correlation matrices for the full update is $VL_t$, and the cost of calculating matrix determinants for the full update is $AV^2L_t$.

\begin{figure}
\begin{centering}
\includegraphics[width=0.6\columnwidth]{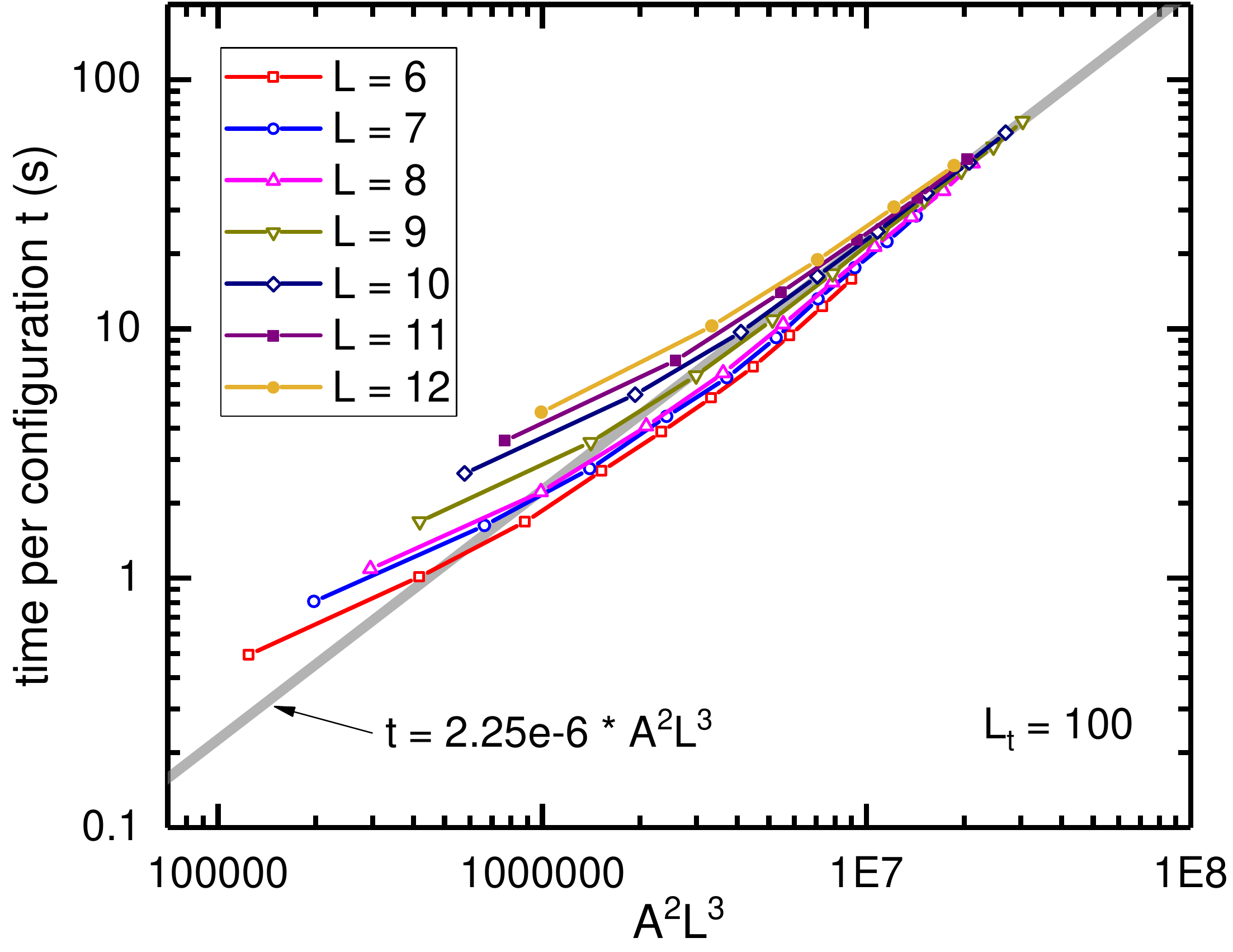}
\par\end{centering}
\caption{\label{fig:timescaling}
    The CPU time for generating a single configuration in units of seconds.
    $A$ and $L$ is the nucleon number and box size, respectively.
    The grey line shows the fitted linear scaling relation for large $A$ or $L$. }
\end{figure}

The PTA has an additional update associated with the pinholes.  The cost of calculating correlation matrices for the full update of all $A$ pinholes is $A^2VL_t$, and the cost of calculating matrix determinants for the full update is $A^4L_t$.
For the values of $A$, $V$, and $L_t$ of interest in this work, the overall computational scaling of the PTA is $A^2VL_t$, while that for the BSS algorithm is $AV^2L_t$.  We see that the cost savings of the PTA is a factor of $V/A$.  We find that the speed up associated with the PTA can be as large as one thousand, depending on the lattice spacing and particle density.

In Fig.~\ref{fig:timescaling} we show the computational time needed to generate one new configuration, consisting of one $\vec{s}$ update and 16 $\vec{c}$ updates.
$A$ is the nucleon number, $L$ is the box size, and
$L_t$ is set to 100.
The grey line shows the fitted linear function, which shows that for large $A$ and $L$ the time scales as $A^2 L^3$.

\subsection{Chemical potential}

Next we discuss the measurement of the observables.
While the energies and density correlation functions can be directly measured by inserting the corresponding operators in the middle time step as in Eq.~(\ref{eq:measurement}), we still need to design efficient algorithms for computing intensive variables, e.g., chemical potential $\mu$ or pressure $p$.
This contrasts with grand-canonical ensemble calculations where the chemical potential is given as an external constraint.

In classical thermodynamics simulations, the Widom insertion method (WIM) \cite{Widom1963} is used to determine the statistical mechanical properties \cite{Binder1997, Dullens2005}.
In the WIM we freeze the motion of the molecules and insert a test particle to the system and measure the free-energy difference, from which the chemical potential can be determined.            
The advantage of the WIM is that we do not need the total free energy, which would require an evaluation of the partition function.
In the PTA we encounter a similar problem.
The absolute free energy can only be inferred with an integration of the energy from absolute zero, which induces large uncertainties.
To solve this problem, we adapt the WIM to the quantum lattice simulations, with the test particles substituted by fermionic particles or holes in the system.



For every configuration $\vec{c}$ generated in the PTA, we calculate the expectation values associated with adding one  nucleon or removing on nucleon.  We define
\begin{eqnarray}
    \mathcal{B}_1 &=& \sum_{c'} \braket{ \vec{c}\cup c'| M(s_{n_t})\cdots M(s_{1}) | \vec{c}\cup c' } / P(\vec{s}, \vec{c}),   \nonumber \\
    \mathcal{B}_{-1} &=& \sum_i \braket{\vec{c} \setminus c_i | M(s_{n_t})\cdots M(s_{1}) | \vec{c}\setminus c_i}/ P(\vec{s}, \vec{c}), \label{eq:widomAmplitudes}
\end{eqnarray}
where the summation over $c'$ runs over all single particle quantum numbers and the summation over $i$ runs over all existing particles.
$P(\vec{s}, \vec{c})$ is the probability given in Eq.~(\ref{eq:PTAprob}).
The extra free energy of inserting or removing one particle is given by
\begin{equation}
    F(A \pm 1) - F(A) = -T \ln \left[\frac{\langle \mathcal{B}_{\pm 1} \rangle_\Omega}{ (A \pm 1)!} \right]   .
\end{equation}
Using the symmetric difference, we have
\begin{equation}
    \mu = [F(A+1) - F(A-1)] / 2 = \frac{T}{2} \ln \left[ A(A+1) \frac{ \langle \mathcal{B}_{-1} \rangle_\Omega } { \langle \mathcal{B}_{1} \rangle_\Omega } \right].
\end{equation}

In the PTA the summations in Eq.~(\ref{eq:widomAmplitudes}) can be calculated using random sampling.
For $\mathcal{B}_1$ we insert a nucleon with random spin and location and propagate it through all time slices, while for $\mathcal{B}_{-1}$ we simply remove one of the existing nucleon.
As only one particle is inserted/removed in each measurement, we find this algorithm very efficient and precise in calculating the chemical potential $\mu$.
Subsequently, we determine the pressure $p$ by integrating the Gibbs-Duhem equation, $dp = \rho d\mu$, starting from the vacuum with $p=0, \rho=0$.

\subsection{Twisted boundary condition}

In any first principles calculation of a quantum many-body system, we are necessarily working with  finite number of nucleons in a finite volume.  The finite volume together with the chosen boundary condition will induce fictitious shell effects.
New lattice magic numbers for protons or neutrons emerge where the calculated observables exhibit unphysical kinks.
It was observed that 66 particles for one species of spin-1/2 fermions give results close to the thermodynamic limit.
This number was extensively used in most of the nuclear matter, neutron matter or cold atom simulations \cite{Forbes2011,Carlson2011}.
However one would ideally like to explore different densities by varying the number of nucleons as well as different the lattice volumes.
For this we must reduce as much as possible the problem of ficitious shell effects.

The origin of the finite volume shell effects is the constraint imposed by the boundary conditions on the particle momenta.
For a cubic box with periodic boundary conditions (PBC), particles are only allowed to have momenta $\bm{p} = \frac{2\pi}{L}\bm{n}$,
which results in a series of magic numbers 2, 14, 38, $\cdots$ for one species of spin-1/2 fermions.
One solution is to use twisted boundary conditions (TBC) \cite{Byers1961} which attach extra phases to wave functions when particles cross the boundaries.
In this case the particle momenta are $\bm{p} = \bm{\theta} + \frac{2\pi}{L}\bm{n}$ for some chosen twist angles $\bm{\theta}$.  It has been found that averaging over all possible twist angles provides an efficient way of approaching the infinite volume limit \cite{Loh1988,Valenti1991,Hagen:2013yba,Schuetrumpf:2016uuk}.

The TBC method was first proposed for exactly solvable models \cite{Loh1988, Valenti1991, Gros1992, Gammel1993, Gros1996} and then found applications in quantum Monte Carlo methods \cite{Lin2001}.
Many groups have applied TBC to lattice QCD calculations to find infinite volume results otherwise not accessible \cite{Bedaque2004, Divitiis2004, Bedaque2005, Sachrajda2004}.
Meanwhile, the application of TBC to lattice effective field theory was shown to be successful, though still limited to few-body and exactly solvable systems \cite{Korber2016}. In this section we discuss the application of the TBC to lattice Monte Carlo calculations and show how it helps remove finite-volume shell effects in thermodynamics calculations.

  We apply the twisted boundary conditions to the single particle wave functions,
  \begin{eqnarray}
      \psi (x + L, y, z, \sigma, \tau) &=& e^{i 2\sigma \theta_x} \psi (x, y, z, \sigma, \tau), \nonumber \\
      \psi (x, y + L, z, \sigma, \tau) &=& e^{i 2\sigma \theta_y} \psi (x, y, z, \sigma, \tau), \nonumber \\
      \psi (x, y, z + L, \sigma, \tau) &=& e^{i 2\sigma \theta_z} \psi (x, y, z, \sigma, \tau),
  \end{eqnarray}
where $-\pi \leq \theta_x$, $\theta_y$, $\theta_z < \pi$ are the independent twist angles in the three directions.
Note that for spin $\sigma = \pm 1/2$ we use the opposite twist angles, which is necessary to preserve time reversal symmetry and avoid sign cancellations.
In this paper we employ TBC by averaging over all possible $(\theta_x, \theta_y, \theta_z),$ which we call average twisted boundary conditions, ATBC.
This can be easily implemented in Monte Carlo calculations by allocating to every thread a random phase triplet with elements uniformly distributed in the interval $[-\pi, \pi)$.

\begin{figure}
\begin{centering}
\includegraphics[width=0.6\columnwidth]{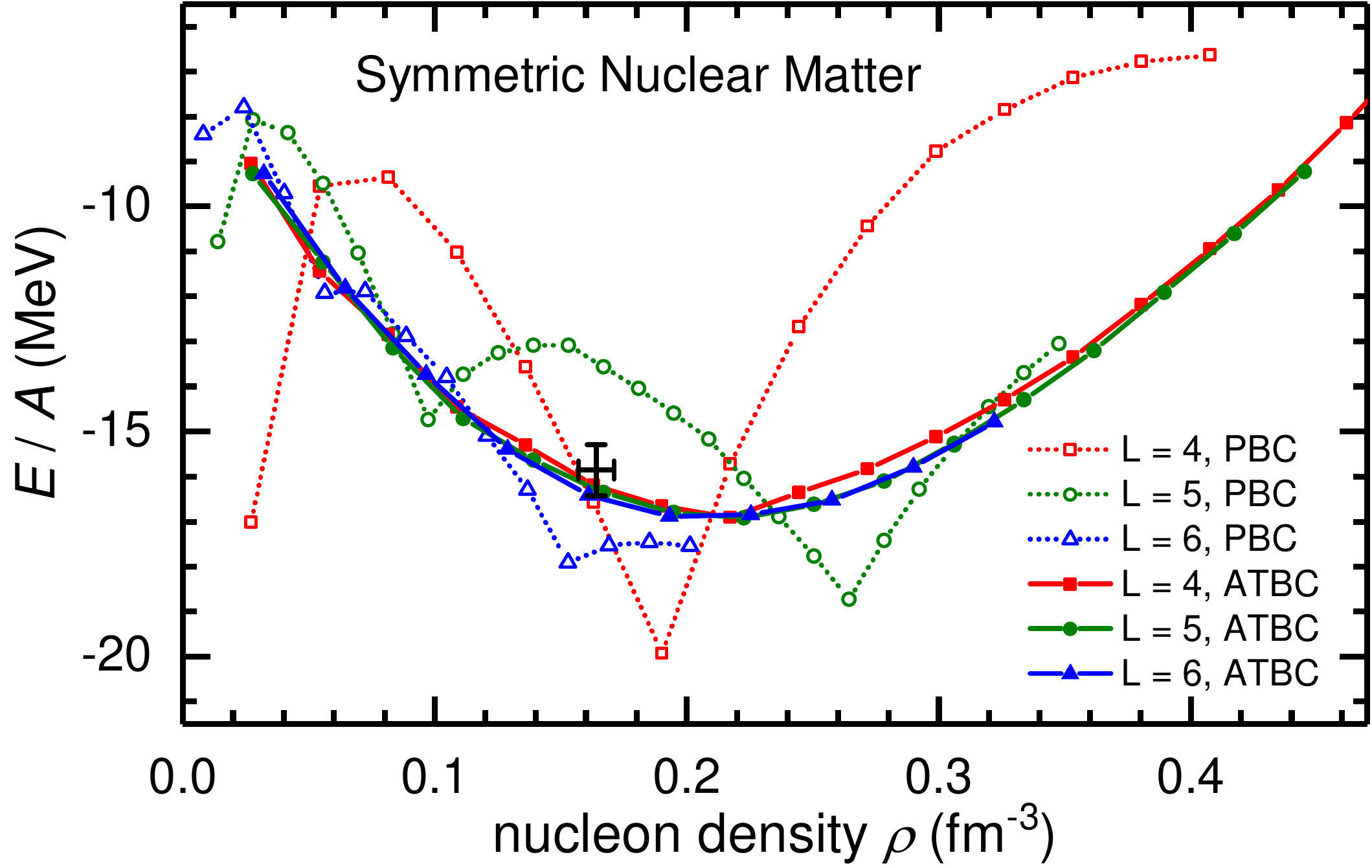}
\includegraphics[width=0.6\columnwidth]{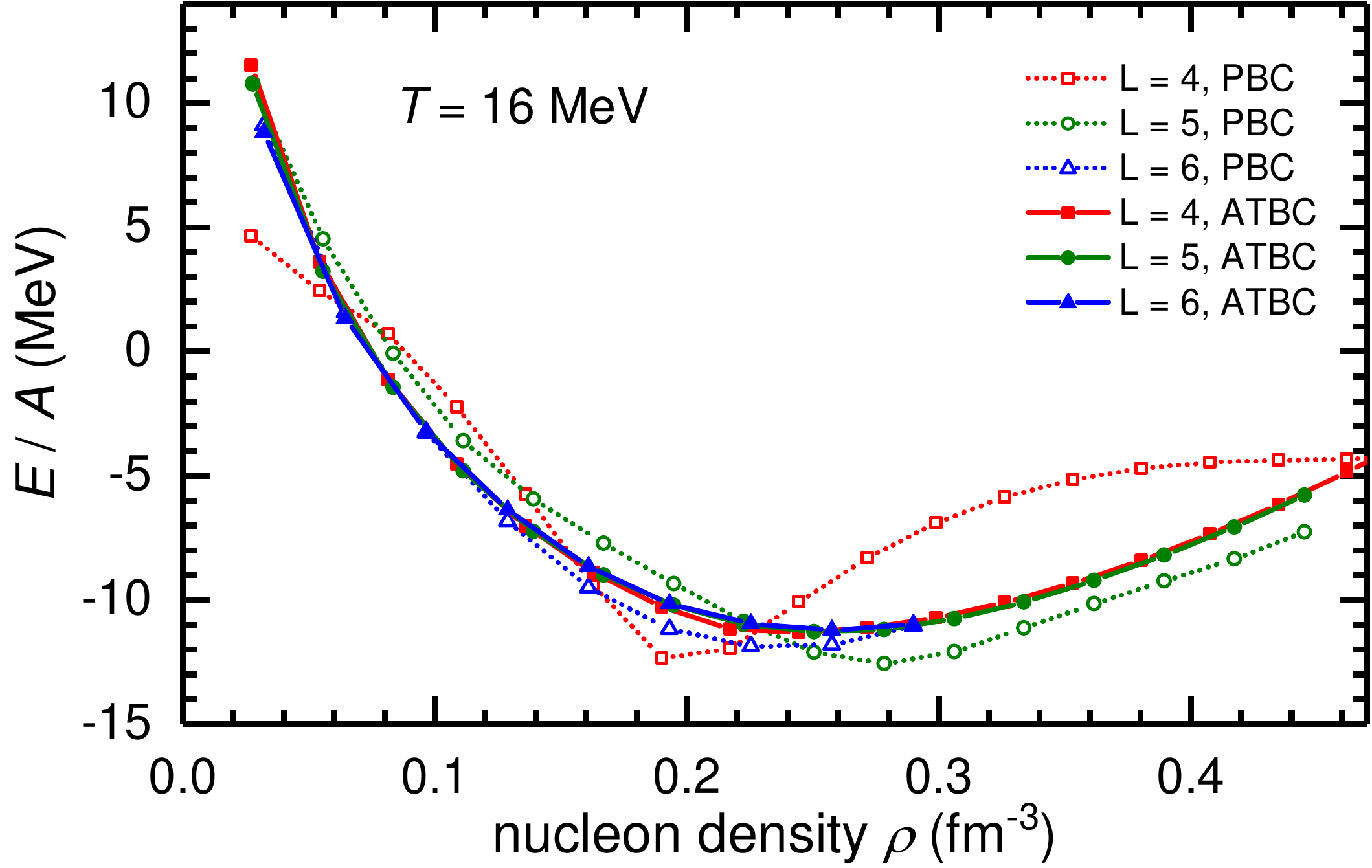}
\par\end{centering}
\caption{\label{fig:ATBC}
(Upper panel) The binding energy per nucleon calculated with periodic boundary conditions (open symbols) and average twisted boundary conditions (full symbols).
    The temperature is $T=0$~MeV.
    The squares, circles and triangles show the results calculated with box sizes $L=$ 4, 5 and 6, respectively.
    The cross with error bars shows the empirical saturation density and binding energy.
    (Lower panel) The binding energy per nucleon at $T=16$~MeV.
    }
\end{figure}

In Fig.~\ref{fig:ATBC} we compare the binding energies at $T=0$ and $T=16$~MeV calculated with different boundary conditions.
For the same density, different box sizes correspond to different nucleon numbers.
The open symbols denote the results calculated with periodic boundary conditions, the full symbols show the results for the average twisted boundary conditions.
Here we see clear shell effects for the PBC calculations.
For each box size the energy oscillates with respect to the nucleon number and exhibit extrema at lattice magic numbers $A=$ 4, 28, 76, $\cdots$.
The amplitudes of the oscillation are smaller for larger boxes, but for $L=6$ the ficitious shell effects are still apparent.
For example, for $T=0$~MeV the energy minimum occurs at $\rho \approx 0.153$~fm$^{-3}$, which is a shell effect that corresponds to $A=76$.
These results can be misleading if we do not take into account the finite volume corrections.
With ATBC, each of the kinks found above disappear and the results collapse onto universal curves.

Some remarks must be added for the finite volume effects.
Here we distinguish between finite volume effects and finite size effects.
The former comes into play together with the boundary conditions and can be removed by using twisted boundary conditions.
However, the latter is due to the finite particle number and manifests itself mainly through the surface effects.
That is, the finite size effects are maximized for inhomogeneous systems, in particular, the system comprising of two or more phases.
The contact surfaces of the different phases give positive contributions to the free energy, which will vanish at the thermodynamic limit.
For example, for symmetric nuclear matter with sub-saturation density, the system we described can be viewed as a large volume of liquid containing a number of small bubbles, with bubble density $\rho=1/V$, where $V$ is the volume used in the simulation.
For large $V$ at fixed density, the bubbles merge together into large ones and the surface effects will eventually disappear.

The finite size effects scale with the surface-volume ratio, which in turn scales as $O(A^{-1/3})$ with respect to the nucleon number.
Thus these effects decay very slowly and cannot be removed with present computational settings.
One example of the surface effects are the upbending of the $T=0$ energy curves at low densities in Fig.~\ref{fig:ATBC}.
For infinite nuclear matter, at the sub-saturation densities the density and the binding energy per nucleon will be exactly the value at the saturation point.
However, for finite systems the extra surface energy causes the upbending of the energy curve and makes it converging to the binding energy per nucleon of small nucleus in the vacuum, $E/A \approx -8$~MeV.


We must stress that the existence of the surface effects at phase coexistence is not a deficiency of our method.
Instead of studying the infinite homogeneous matter, our method focuses on the real finite systems with phenomena like cluster and phase separation.
Consequently, we believe that our formalism, together with the advanced nuclear interactions, will pave the way of fully understanding the nuclear thermodynamic processes.

\section{Results}

In this work we perform simulations on a $6\times 6 \times 6$ cubic lattice.
The temporal lattice spacing is $a_{t}=1/2000$~MeV$^{-1}$.
We impose twisted boundary conditions along the $x$-, $y$- and $z$- directions~\cite{Lin2001}.
The twist angles are averaged over all possible values by Monte Carlo sampling to remove the fictitious finite-volume shell effect.

\begin{figure}
\begin{centering}
\includegraphics[width=0.6\columnwidth]{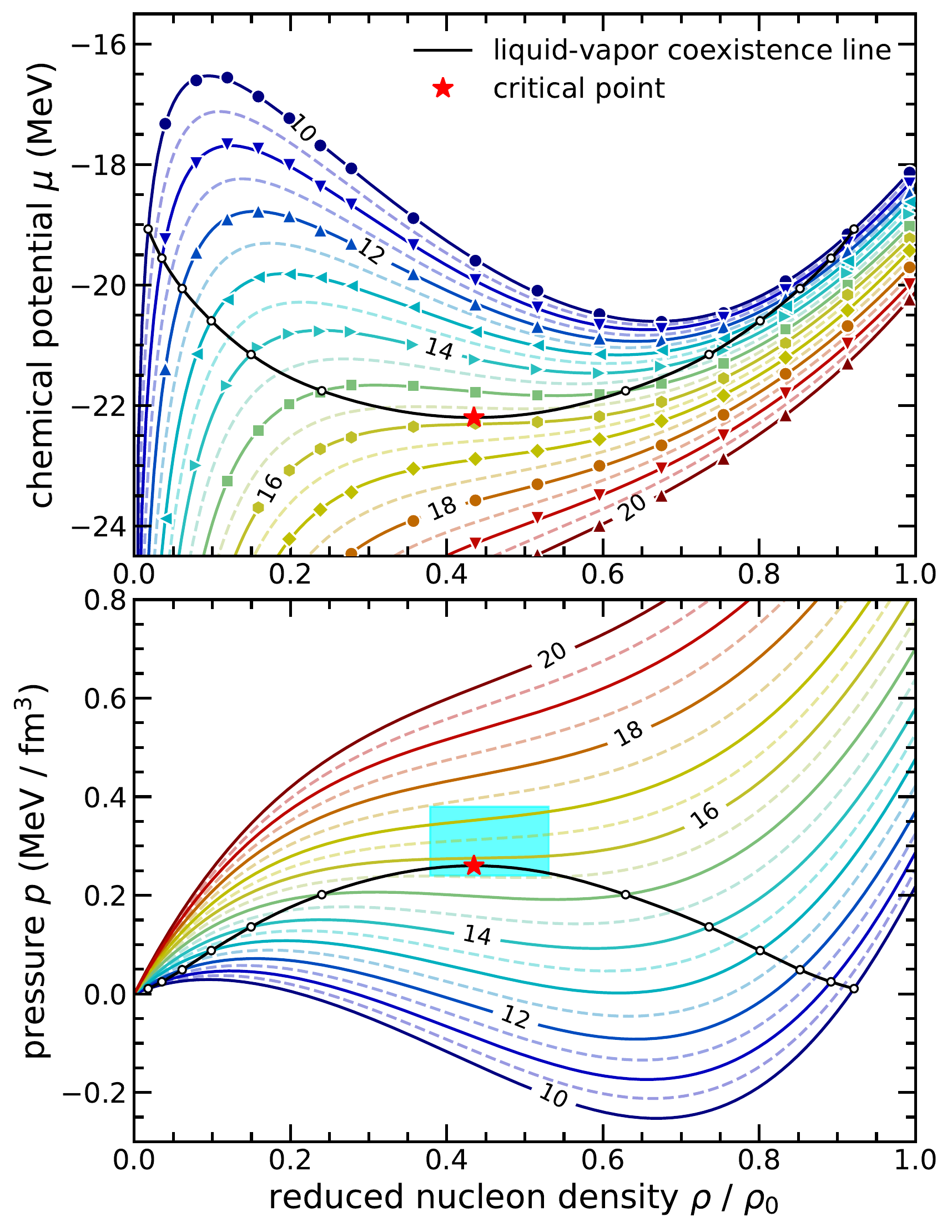}
\par\end{centering}
\caption{\label{fig:isotherms}
(Upper panel) The $\mu$-$\rho$ isotherms of symmetric nuclear matter. The nucleon densities are re-scaled against the saturation density $\rho_{0}$. The symbols represent the lattice
results, the error bars are smaller than the symbols. The connecting lines are interpolations. The numbers are the corresponding temperatures.
The temperature differences between adjacent solid isotherms are 1~MeV. The black line denotes the liquid-vapor coexistence line
derived from the Maxwell construction. The red star denotes the calculated critical point.
    (Lower panel) The $p$-$\rho$ isotherms. The cyan square marks the empirical critical point extracted from experiment~\cite{Elliott2013}.
}
\end{figure}

In this work we study the nuclear liquid-vapor phase transition by examining the finite-temperature equation of state.
Throughout, we only consider symmetric nuclear matter with equal numbers of protons and neutrons and the Coulomb interaction is neglected.
In Fig.~\ref{fig:isotherms} we present the calculated chemical potential and pressure isotherms.
Each point represents a separate simulation.
The temperature $T$ covers the range of 10.0~MeV $\leq T \leq$ 20.0~MeV and the nucleon number $A$ varies from 4 to 100,
which corresponds to densities from 0.008~fm$^{-3}$ to 0.2~fm$^{-3}$.
These settings allow us to explore the whole region relevant to the nuclear matter liquid-vapor phase transition.
We found that the quantum Widom insertion method (QWIM) gives highly precise measurements of the chemical potential $\mu$ for a large range of densities from a dilute nucleon gas to supersaturation density 2$\rho_0$ over the entire temperature regime considered in this paper.
All the Monte Carlo errors for $\mu$ are smaller than 0.02~MeV and not shown explicitly in Fig.~\ref{fig:isotherms}.
Based on the lattice results, we map the whole $\mu$-$\rho$-$T$ equation of state in this area using interpolation.
The critical point is then deduced from solving the equations $d\mu/d\rho = d^2\mu/d\rho^2 = 0$.
The uncertainties in the critical values are estimated by propagating the simulation and interpolation errors.
We found the critical temperature, density, and chemical potential to be $T_c=$15.80(3)~MeV, $\rho_c=$0.089(1)~fm$^{-3}$, and $\mu_c$= $-22.20$(1) MeV, respectively.
The liquid-vapor coexistence line is determined through the Maxwell construction of each isotherm and depicted as a solid black line in Fig.~\ref{fig:isotherms}.

All the other bulk thermodynamic quantities can be reliably extracted based on the calculated high-precision energies and chemical potentials.
In the lower panel of Fig.~\ref{fig:isotherms} we show the deduced pressure-density isotherms, the corresponding liquid-vapor coexistence line and the critical point.
The calculated critical pressure is $P_c=$0.260(3)~MeV/fm$^{3}$.
For comparison, we also draw the critical point extracted by analyzing the cluster distributions in heavy-ion collisions,
$T_c=17.9(4)$~MeV, $\rho_c=0.06(1)$~fm$^{-3}$ and $P_c=0.31(7)$~MeV/fm$^{3}$~\cite{Elliott2013}.
Note that our calculations employ a leading-order chiral interaction which also overestimates the nuclear matter saturation density $\rho_0$,
we expect that the quality of both $\rho_0$ and $\rho_c$ calculations can be improved by including higher-order corrections.
These corrections should all be perturbative and not change the essence of the physics discussed in this paper.

\begin{figure}
\begin{centering}
\includegraphics[width=0.6\columnwidth]{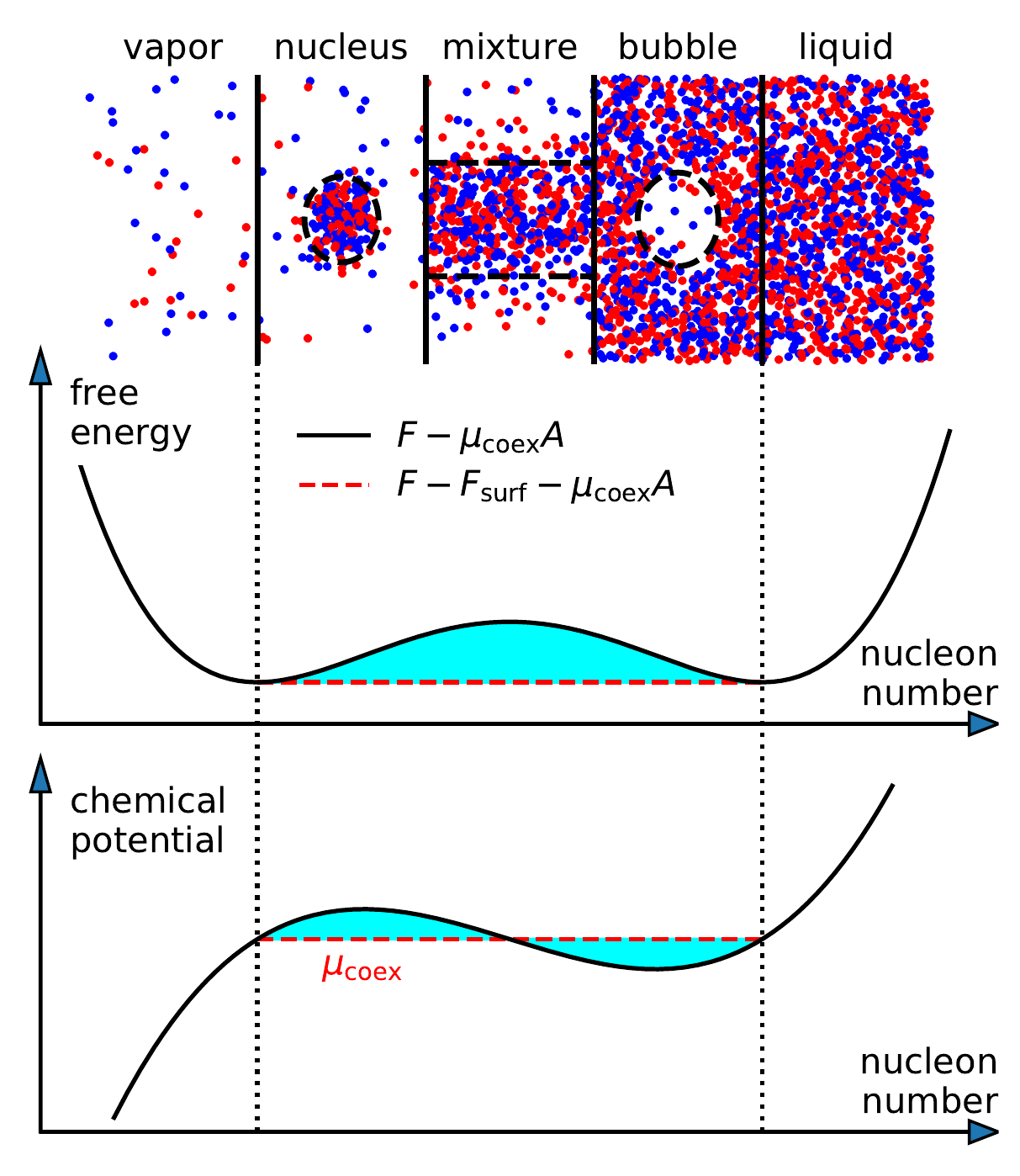}
\par\end{centering}
\caption{\label{fig:schematic}
    The schematic plots of the vapor-liquid phase transition in a finite nuclear system with fixed volume $V$ and temperature $T<T_c$.
    (Upper panel) The most probable configurations for different nucleon number $A$.
    Red (blue) points stand for protons (neutrons).
    The dashed lines signify the surfaces seperating the liquid and vapor phases.
    (Middle panel) The free energy.
    Solid (dashed) lines denote the results with (without) the surface contributions.
    (Lower panel) The chemcial potential $\mu = \partial{F}/\partial{A}$.
}
\end{figure}

The calculated isotherms follow exactly the pattern expected for a liquid-vapor phase transition in a finite system.
Above $T_c$ the system is in a supercritical state, while below $T_c$ the pure liquid and vapor phases exist in the high- and low-density regime, respectively.
For states encompassed by the two arms of the coexistence line, the system is a mixture of the liquid and vapor phases.
In the thermodynamic limit, where $N, V \rightarrow \infty$ with $N/V$ kept finite, $\mu$ and $p$ are constants in the coexistence regime along an isotherm,
both of which are uniquely determined by the chemical and mechanical equilibrium conditions, $\mu_l = \mu_v = \mu_{\rm coex}$ and $p_l = p_v = p_{\rm coex}$,
where the subscripts $l$ and $v$ denote the liquid and vapor phases, respectively.
For a finite system the above conditions still hold; however, the surface effects are usually non-neglibible and $\mu_{\rm coex}$ and $p_{\rm coex}$ can have different values.
A well-known example is that the pressure of the vapor in equilibrium with small liquid drops can be larger than its thermodynamic-limit value,
with the difference compensated by the contribution of the surface tension.
Bearing the importance of the surface contributions in mind, we can easily interpret the \textit{ab initio} calculations presented in Fig.~\ref{fig:isotherms}.

The most prominent feature of the isotherms in Fig.~\ref{fig:isotherms} is the backbending in the coexistence regime below $T_c$.
Note that the origin of this backbending is completely different from that of similar structures found in the van der Waals model or other mean-field calculations.
The mean-field models always describe homogeneous systems and the backbending of the $p$-$\rho$ isotherms result in a negative compressibility,
and in this regard the assumption of homogeneity conflicts with the condition of mechanical equilibrium.
Conversely, in \textit{ab initio} calculations we do not rely on the assumption of homogeneity; the results always describe realistic systems.
In particular, phase separation occurs spontaneously whenever it is favored by the free-energy criterion.
In the coexistence regime, the most probable configurations consist of high-density liquid regions and low-density vapor regions, the surface spatially separating these regions
gives rise to a positive contribution to the total free energy, which prohibits the formation of small liquid drop in vapor or small bubbles in liquid.
The distortions of the isotherms reflect the efforts of the system to overcome such a surface-energy barrier.

In Fig.~\ref{fig:schematic} we show schematic plots illustrating the underlying mechanism.
Given a fixed volume $V$ and a temperature $T$ below the critical value $T_c$, the free energy $F$ is a function of the nucleon number $A$.
In the middle panel of Fig.~\ref{fig:schematic} we show the free energy curve across the liquid-vapor coexistence region.
We subtract $\mu_{\rm coex} A$ from $F$ to remove most of the $A$-dependence, with $\mu_{\rm coex}$ assuming the value at the thermodynamic limit.
For a finite system the surface free energy $F_{\rm surf}$ is approximately proportional to the area of the surface.
In the upper panel of Fig.~\ref{fig:schematic} we show the most probable configurations for different densities.
At low densities we have a nucleus surrounded by small clusters, while at high densities we see bubbles in a nuclear liquid.
At intermediate densities the system contains bulk nuclear matter with appreciable surface areas.
The surface area first increases after the formation of a nucleus then decreases when most of the volume is occupied by the liquid phase.
Correspondingly $F_{\rm surf}$ has a unique maximum and creates a bump in the free energy curve.
In the lower panel of Fig.~\ref{fig:schematic} we show the corresponding chemical potential $\mu = \partial{F}/\partial{A}$.
Apparently the backbending is a natural result of the surface free energy contributions.

\begin{figure}
\begin{centering}
\includegraphics[width=0.6\columnwidth]{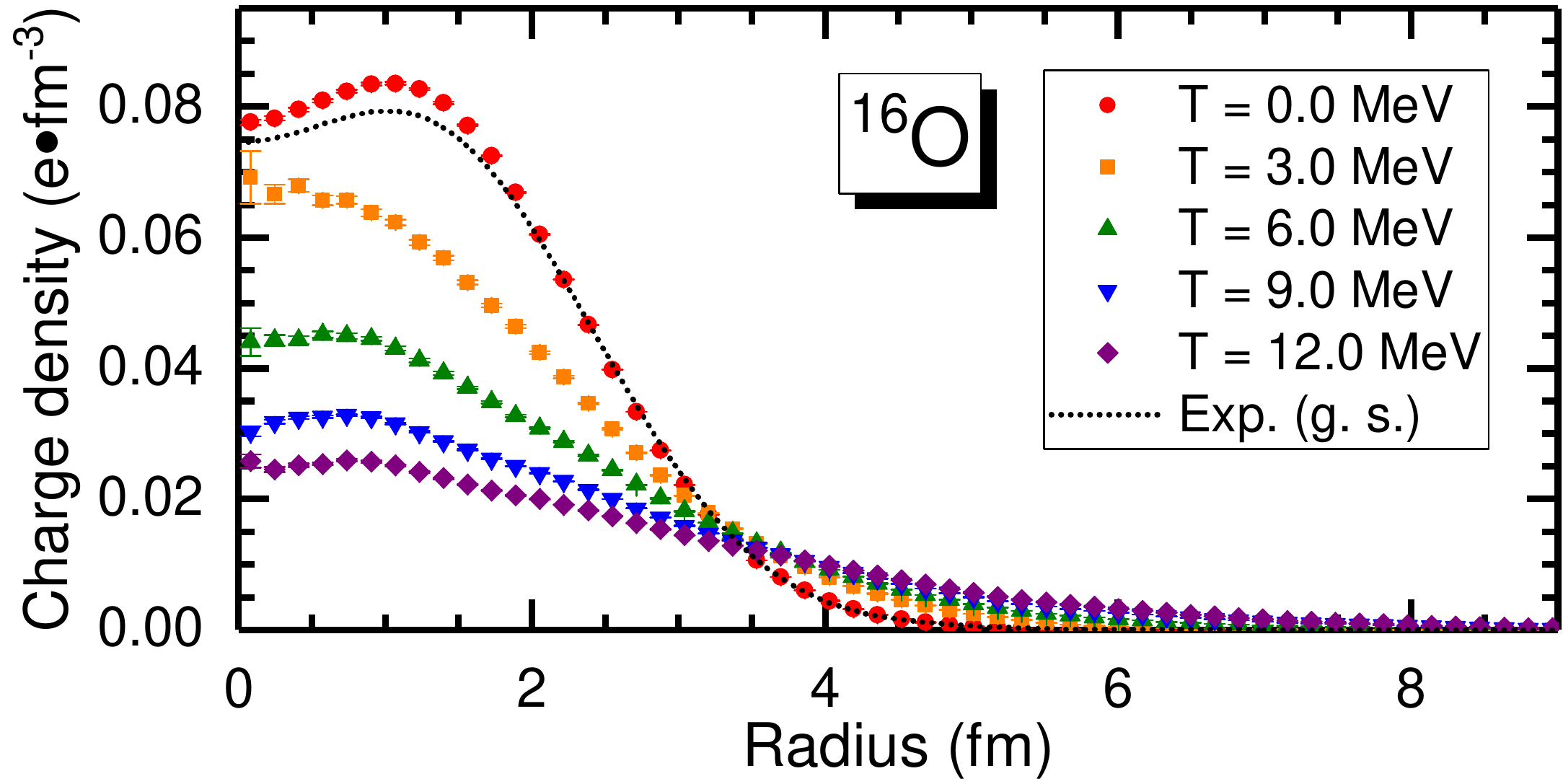}
\par\end{centering}
\caption{\label{fig:O16density}
The charge density distribution of $^{16}$O calculated in a $L=8$ box.
The circles (red), squares (orange), up triangles (green), down triangles (blue) and diamonds (purple) denote the results calculated with temperature $T=0, 3, 6, 9$ and $12$ MeV, respectively.
The dotted line denotes the experimental values.
    }
\end{figure}

Another important scenario where the nuclear temperature play a key role is the highly excited states of the finite nuclei.
These states can be accessed through nuclear collisions with high bombarding energies.
In contrast to the compact ground state, in these energetic states the nucleons can move freely in a much larger volume.
The corresponding huge phase space can only be described with the concepts from the thermodynamics.
In Fig.~\ref{fig:O16density} we show the charge density profiles of $^{16}$O calculated at different temperatures.
At $T=0$ MeV the experimental ground state density shown by dotted line are well reproduced by our Hamiltonian.
For higher temperatures the charge density becomes more diffuse, signifying stronger coupling with the continuum states.
When $T$ is large enough, firstly the $\alpha$-clusters lastly the individual nucleons will be discharged from the nucleus and move freely as a nuclear vapor.
However, even for $T=12$ MeV the density still accumulates near the origin $r=0$ fm, which is a consequence of the density correlations induced by the attractive nuclear force.
Note that the densities in Fig.~\ref{fig:O16density} are calculated with the single-nucleon density operators.
The many-body correlations contain more interesting information that can not be described the mean field theories.
Exploring these aspects using {\it ab initio} methods such as the PTA presented in this work can be a promising direction in the near future.

\section{Summary and perspective}

In this work we have presented the details of the pinhole trace algorithm
for simulating a nucleus at fixed temperature $T$, volume
$V$ and particle number $A$. We trace over the $A$-body Hilbert space using
auxiliary-field Monte Carlo method and find that the Monte Carlo sign oscillations
are under control.  Because we are working with the canonical ensemble, it suffices
to propagate in imaginary time only $A$ single-nucleon states, in contrast with the grand canonical
ensemble where the full $4L^3$ space of single-nucleon states must be propagated.
This provides an enormous computational advantage over grand canonical ensemble
simulations that can be a factor of several thousands to as much as
several millions for large volume simulations.

Compared with shell-model based methods, the {\it ab initio lattice} formalism
enables us to explore a much larger configuration space, which is essential
for studying phenomena involving high excitation energies and strong
many-body correlations.
This work is only a first exploration into {\it ab initio} nuclear thermodynamics with realistic interaction; there is plenty of more work to be done.
When combined with high-accuracy lattice chiral interactions
that are improved systematically order by order \cite{Li2018}, we can perhaps finally realize the goal of
\textit{ab initio} calculations with controlled systematic errors that explore all aspects of the
nuclear equation of state as a function of density, temperature, and proton fraction.

\section*{Acknowledgments}

We are grateful for discussions with Pawel Danielewicz.
We acknowledge partial financial support from
the Deutsche Forschungsgemeinschaft (TRR 110,
``Symmetries and the Emergence of Structure in QCD\textquotedblright ),
the BMBF (Verbundprojekt
05P18PCFP1), the U.S. Department of Energy (DE-SC0013365 and DE-SC0021152),
the National Science Foundation (grant no. PHY1452635), and the Scientific
and Technological Research Council of Turkey (TUBITAK project no.
116F400). Further support was provided by NSAF
(Grant No. U1930403), the Chinese Academy of Sciences
(CAS) President\textquoteright s International Fellowship Initiative
(PIFI) (grant no. 2018DM0034) and by VolkswagenStiftung (grant no.
93562). This project has received funding from the European Research Council (ERC) under the
European Union's Horizon 2020 research and innovation programme (grant agreements No. 101018170 and 885150).
The computational resources were provided by the Gauss
Centre for Supercomputing e.V. (www.gauss-centre.eu) for computing time on the GCS Supercomputer JUWELS at J{\"u}lich
Supercomputing Centre (JSC), Oak Ridge Leadership Computing
Facility, RWTH Aachen, Michigan
State University and the Beijing Super Cloud Computing Center (BSCC, http://www.blsc.cn/).



\begin{thebibliography}{99}
\bibitem{Togashi2017}H.~Togashi, K.~Nakazato, Y.~Takehara, S.~Yamamuro, H.~Suzuki, M.~Takanobe, Nucl. Phys. A
961, 78 (2017).

\bibitem{Page2004}D.~Page, J.~M.~Lattimer, M.~Prakash, A.~W.~Steiner, Astrophys. J. Suppl. 155, 623 (2004).

\bibitem{Most2019}E.~R.~Most, L.~J.~Papenfort, V.~Dexheimer, M.~Hanauske, S.~Schramm, H.~Stöcker, L.~Rezzolla, Phys. Rev. Lett. 122, 061101 (2019).

\bibitem{Hauser1952}W. Hauser and H. Feshbach, Phys. Rev. 87, 366 (1952).

\bibitem{Bohr1936}N. Bohr, Nature 137, 351 (1936).

\bibitem{Siemens1983}P. J. Siemens, Nature 305, 410 (1983).

\bibitem{Seeger1965}P. Seeger, W. Fowler, and D. Clayton, The Astrophysical
Journal (1965).

\bibitem{Freer2018}M. Freer, H. Horiuchi, Y. Kanada-En\textquoteright yo,
D. Lee, and U.-G. Mei{\ss}ner, Rev. Mod. Phys. 90, 035004 (2018).

\bibitem{Koonin1997}S.E. Koonin, D.J. Dean, and K. Langanke, Phys.
Rep. 278, 1 (1997)

\bibitem{Dean1995}D. J. Dean, S. E. Koonin, K. Langanke, P. B. Radha,
and Y. Alhassid, Phys. Rev. Lett. 74, 2909 (1995).

\bibitem{Nakada1997}H. Nakada and Y. Alhassid, Phys. Rev. Lett. 79,
2939 (1997).

\bibitem{Alhassid2000}Y. Alhassid, G. F. Bertsch, S. Liu, and H.
Nakada, Phys. Rev. Lett. 84, 4 (2000).

\bibitem{Aichelin1991}J. Aichelin, Phys. Rep. 202, 233 (1991).

\bibitem{Ono1992b}A. Ono, H. Horiuchi, T. Maruyama, A. Ohnishi, Phys.
Rev. Lett. 68, 2898 (1992).

\bibitem{Fisher1967}M.E. Fisher, Physics-New York 3, 255 (1967).

\bibitem{Fisher1967b}M.E. Fisher, Rep. Prog. Phys. 30, 615 (1967).

\bibitem{Furuta2006}T. Furuta and A. Ono, Phys. Rev. C 74, 014612 (2006)

\bibitem{Ono2019}A. Ono, Prog. Part. Nucl. Phys. 105, 139 (2019).


\bibitem{Epelbaum2009}E. Epelbaum, H.-W. Hammer, and U.-G. Mei{\ss}ner,
Rev. Mod. Phys. 81, 1773 (2009).

\bibitem{Muller1999}H.~M.~M{\"u}ller, S.~E.~Koonin, R.~Seki, U.~van~Kolck, Phys. Rev. C 61, 044320 (2000).

\bibitem{Lee2004}D.~Lee, B.~Borasoy, T.~Sch{\"a}fer, Phys. Rev. C 70, 014007 (2004).


\bibitem{Lu2020}Bing-Nan Lu, Ning Li, Serdar Elhatisari, Dean Lee, Joaqu{\'i}n~E.~Drut, Timo A. L{\"a}hde, Evgeny Epelbaum and Ulf-G. Mei{\ss}ner, Phys. Rev. Lett. 125, 192502 (2020)

\bibitem{Lu2018}B.-N. Lu, N. Li, S. Elhatisari, D. Lee,
E. Epelbaum, U.-G. Mei{\ss}ner, Phys. Lett. B 797, 134863 (2019).

\bibitem{Lee2009}D. Lee, Prog. Part. Nucl. Phys. 63, 117 (2009).

\bibitem{Lahde2019}T.~A.~L{\"a}hde, U.-G.~Mei{\ss}ner, ``Nuclear Lattice Effective Field Theory: An Introduction'', Lecture Notes in Physics, Volume 957, Springer, (2019).

\bibitem{Elhatisari2017}S. Elhatisari, E. Epelbaum, H. Krebs, T.
A. L{\"a}hde, D. Lee, N. Li, B. Lu, U.-G. Mei{\ss}ner, and G. Rupak, Phys.
Rev. Lett. 119, 222505 (2017).


\bibitem{Blankenbecler:1981jt}
R.~Blankenbecler, D.~J.~Scalapino and R.~L.~Sugar,
Phys. Rev. D \textbf{24}, 2278 (1981),

\bibitem{Widom1963} B. Widom, J. Chem. Phys. 39, 2808 (1963).

\bibitem{Binder1997} K. Binder, Rep. Prog. Phys. 60, 487 (1997).

\bibitem{Dullens2005} R.P.A. Dullens, Mol. Phys. 103, 3195 (2005).

\bibitem{Forbes2011} M. Forbes, S. Gandolfi, and A. Gezerlis, Phys. Rev.
Lett. {\bf 106}, 235303 (2011).

\bibitem{Carlson2011} J. Carlson, Sefano Gandolfi, Kevin E. Schmidt, Shiwei
Zhang, Phys. Rev. A {\bf 84}, 061602R (2011).

\bibitem{Byers1961} N. Byers and C. N. Yang, Phys. Rev. Lett. {\bf 7}, 46
(1961).

\bibitem{Loh1988} E. Y. Loh, Jr. and D. K. Campbell, Synthetic Metals {\bf
27},
A499 (1988).

\bibitem{Hagen:2013yba}
  G.~Hagen, T.~Papenbrock, A.~Ekström, K.~A.~Wendt, G.~Baardsen, S.~Gandolfi,
M.~Hjorth-Jensen and C.~J.~Horowitz,
  Phys.\ Rev.\ C {\bf 89}, no. 1, 014319 (2014).

\bibitem{Schuetrumpf:2016uuk}
  B.~Schuetrumpf, W.~Nazarewicz and P.-G.~Reinhard,
  Phys.\ Rev.\ C {\bf 93}, no. 5, 054304 (2016).

\bibitem{Valenti1991} R. Valenti, C. Gros, P. J. Hirschfeld and W. Stephan,
Phys. Rev. B {\bf 44}, 13203 (1991).

\bibitem{Gros1992} C. Gros, Z. Phys. B {\bf 86}, 359 (1992).

\bibitem{Gammel1993} J. Tinka Gammel, D. K. Campbell, and E. Y. Loh, Jr.,
Synthetic Metals {\bf 55}, 4437 (1993).


\bibitem{Gros1996} C. Gros, Phys. Rev, B {\bf 53}, 6865 (1996).


\bibitem{Lin2001} C. Lin, F.-H. Zong, D. M. Ceperley, Phys. Rev. E {\bf 64},
016702
(2001).


\bibitem{Bedaque2004} Paulo F. Bedaque, Phys. Lett. B {\bf 593}, 82 (2004).

\bibitem{Divitiis2004} G. M. Divitiis, R. Petronzio, N. Tantalo, Phys. Lett.
B {\bf 595}, 408 (2004).

\bibitem{Bedaque2005} Paulo F. Bedaque, Jiunn-Wei Chen, Phys. Lett. B {\bf
616}, 208 (2005).

\bibitem{Sachrajda2004} C. T. Sachrajda, G. Villadoro, Phys. Lett. B {\bf
609}, 73 (2005).

\bibitem{Korber2016} C. K{\"o}rber, T. Luu, Phys. Rev. C {\bf 93}, 054002
(2016).

\bibitem{Elliott2013} J. B. Elliott, P. T. Lake, L. G. Moretto, L. Phair, Phys. Rev. C 87, 054622 (2013).



%
%
%
%
%
%
%
%
\bibitem{Li2018}N. Li, S. Elhatisari, E. Epelbaum, D. Lee, B.-N.
Lu, and U.-G. Mei{\ss}ner, Phys. Rev. C 98, 044002 (2018).
%
%
%
%
%
%
%
%
%
%
%
%
%
%
%
%
%
%
%
%
%
%
%
%
%
%
%
%
%
%
%
%
%
%
%
%
%
%
%
%
%
%
%
%
%
%
%
%
%
%
%
%
%
%
%
%
%
%
%
%
%
%
%
%
%
%
%
%
%
%
%
%
%
%
%
%
%
%




\end{thebibliography}
\end{document}